\documentclass[intlimits,twoside,a4paper]{article}

\usepackage{amsmath,amssymb}
\usepackage{graphicx}

\usepackage{color}

\usepackage[T2A]{fontenc}
\usepackage[cp1251]{inputenc}
\usepackage{cmpj2}

\issue{2013}{16}{1}{14002}

\doinumber{10.5488/CMP.16.14002}

\articletype{Rapid Communication}

\title[Nematic fluid at a hard wall in the mean field approximation]%
{Nematic fluid at a hard wall in the mean field approximation}%

\author[M. Holovko, I. Kravtsiv, D. di Caprio]{ M. Holovko\refaddr{label1}, I. Kravtsiv\refaddr{label1}\thanks{E-mail: ivankr@icmp.lviv.ua}\ , D. di Caprio\refaddr{label2}
}
\addresses{
\addr{label1} Institute for Condensed Matter Physics of the National Academy of Sciences of Ukraine, \\
1 Svientsitskii St., 79011 Lviv, Ukraine
\addr{label2} Laboratoire d'Electrochimie, Chimie des Interfaces et
Mod\'elisation pour l'Energie (LECIME) ENSCP, \\ Chimie ParisTech, Case 39, 4, Pl.
Jussieu, 75005 Paris, France.}

\date{Received January 31, 2013, in final form February 14, 2013}
\authorcopyright{M. Holovko, I. Kravtsiv, D. di Caprio, 2013}

\begin{document}

\maketitle

\begin{abstract}

In the framework of a field theoretical approach we study
Maier-Saupe nematogenic fluid in contact with a hard wall. The
pair interaction potential of the considered model consists of
an isotropic and an anisotropic Yukawa terms. In the mean field
approximation the contact theorem is proved. For the case of
the nematic director being oriented perpendicular to the wall,
analytical expressions for the density and order parameter
profiles are obtained. It is shown that in a certain
thermodynamic region the nematic fluid near the interface can
be more diluted and less orientationally ordered than in the
bulk region.



\end{abstract}

Due to orientational ordering, nematic fluids near a surface
show richer behavior than in the case of simple fluids. Among
them are the anchoring phenomena, whereby the surface induces a
specific orientation of the nematic director with respect to
the surface \cite{jerome}. In order to understand this
phenomenon, during the past decade the Henderson-Abraham-Barker
(HAB) approach \cite{sokol1,sokol2}, previously developed in
the theory of isotropic fluids in contact with solid surfaces
\cite{hend}, has been employed. In this approach, the description
of the fluid density profile reduces to the solution of the
Ornstein-Zernike (OZ) integral equation for the fluid
particle-wall distribution function calculated from the known
fluid particle distribution function in the bulk. In the framework of the HAB
approach, the application to the bulk of the nematic model, analytically
solvable at the level of the mean spherical approximation (MSA)~\cite{holsok},
makes it possible  to investigate the role of orientational-dependent
molecular interactions with the surface in anchoring phenomena
\cite{sokol1,sokol2}. However, in the MSA, this approach does
not take into account the contribution from long-range
molecular interactions and, as a result, does not satisfy the
exact relation known as the contact theorem
\cite{contactth1,contactth2}. According to this theorem, the contact value of the particle density near a
hard wall for a neutral fluid  is determined by the pressure of the fluid in the
bulk volume.

Recently, the density field theory, previously developed for
ionic fluids near a hard wall \cite{ionic1,castaba1,castaba2},
has been applied to the description of simple fluids with
Yukawa-type interactions near a hard wall \cite{molphys}. In
both cases, the developed approach yielded correct results. In this
theory, the contributions from the mean field and from the
fluctuations are separated. In \cite{molphys}, it was shown that
the mean field treatment of a Yukawa fluid near a wall reduces
to solving a non-linear differential equation for the density
profile while the treatment of fluctuations reduces to the OZ
equation with the Riemann boundary condition.

In this paper, the density field theory, developed in
\cite{molphys} for simple fluids at a hard wall, will be
generalized to nematic fluids at a hard wall.
To this end, we consider Maier-Saupe
nematogenic fluid model \cite{masa,masa_2} as one of the simplest
models that account for the isotropic-nematic phase transition.
For simplification we consider a fluid of point uniaxial
nematogens interacting through the pair potential
\begin{align}
\nu(r_{12},\Omega_1\Omega_2)=\nu_0(r_{12})+\nu_2(r_{12})P_2(\cos\theta_{12}),
\label{potential}
\end{align}
where the first term
$\nu_0(r_{12})=\left({A_0}/{r_{12}}\right)\exp\left({-\alpha_0
r_{12}}\right)$ describes the isotropic repulsion and the
second term with
$\nu_2(r_{12})=\left({A_2}/{r_{12}}\right)\,\exp\left({-\alpha_2
r_{12}}\right)$ describes the anisotropic attraction between
particles ($A_0>0$, $A_2<0$), $r_{12}$ denotes the distance
between particles $1$ and $2$, $\Omega=\left(\theta,\phi\right
)$ are orientations of particles,
$P_2(\cos{\theta_{12}})=(3\cos^2\theta_{12}-1)/2$ is the second
order Legendre polynomial of the relative orientation
$\theta_{12}$.

The application of the density field theory to the description
of bulk properties of such nematic fluids was considered in
\cite{cond,kravtsiv}. It was shown that beyond the mean field
approximation, the repulsive isotropic term in (\ref{potential})
is very important for the description of the nematic phase.
Within the field-theoretical formalism, the Hamiltonian is a
functional of the density field and can be written as the sum
of entropic and interaction terms
    \begin{align}
    \beta H[\rho(\mathbf{r},\Omega)]=&\int\rho({\mathbf{r}},\Omega)\left[\ln(\rho(\mathbf{r},\Omega)
    \Lambda_R\Lambda_T^3)-1\right]\rd{\mathbf{r}\rd\Omega}\\
    &+\frac{\beta}{2}\int{\nu(r_{12},\Omega_1\Omega_2)\rho(\mathbf{r_1},\Omega_1)\rho(\mathbf{r_2},\Omega_2)}
    \rd{\mathbf{r_1}}\rd{\mathbf{r_2}\rd{\Omega_1}\rd{\Omega_2}}\, ,\nonumber
    \end{align}
where $\beta=1/k_\mathrm{B}T$ is the inverse temperature,
$\rd\Omega=(1/4\pi)\sin\theta \rd\theta \rd\phi$ is the normalized
angle element, $\rho(\mathbf{r},\Omega)$ is particle density
per angle, so that
$\int{\rho(\mathbf{r},\Omega)\rd{\Omega}}=\rho(\mathbf{r})$,
$\Lambda_T$ is the thermal de Broglie wavelength of the
molecules, the quantity $\Lambda^{-1}_R$ is the rotational
partition function for a single molecule \cite{gra}.

In this paper, we will restrict our consideration to the mean
field approximation (MFA) which is the lowest order
approximation for the partition function. In the canonical
formalism it corresponds to fixing the Lagrange parameter
$\lambda$, so that the following relation is true for the
singlet distribution function
\begin{equation}
\label{MFA1}
\frac{\delta\beta
H[\rho(\mathbf{r},\Omega)]}{\delta\rho(\mathbf{r},\Omega)}\Bigg\vert_{\rho^\mathrm{MFA}}=\lambda.
\end{equation}
As a result,
\begin{align}
\label{MFA_s}
\rho(\mathbf{r}_1,\Omega_1)=\rho^\mathrm{bulk}(\Omega_1)\exp\left\{-\beta\int\nu(r_{12},\Omega_1\Omega_2)
\left[\rho(\mathbf{r}_2,\Omega_2)-
\rho^\mathrm{bulk}(\Omega_2)\right]\rd\mathbf{r_2}\rd{\Omega_2}\right\},
\end{align}
where
\begin{align}
\rho^\mathrm{bulk}(\Omega)=\rho_\mathrm{b}{\exp\left[-\left(\varkappa_2^2S_\mathrm{b}/\alpha_2^2\right)
\,P_2(\cos\theta)\right]}\Big/
{\int\limits_0^1 \rd\cos\theta\exp\left[-\left(\varkappa_2^2S_\mathrm{b}/\alpha_2^2\right)\,P_2(\cos\theta)\right]}
\end{align}
is the singlet distribution function for the bulk nematic in
the MFA defined in the framework of the Maier-Saupe theory
\cite{cond,kravtsiv}, $\varkappa_2^2=4\pi\rho_\mathrm{b}\beta A_2$,
$\rho_\mathrm{b}$ is the bulk value of the fluid density,
$S_\mathrm{b}=(1/\rho_\mathrm{b})\int_0^1
P_2(\cos\theta)\rho^\mathrm{bulk}(\Omega)\rd\cos\theta$ is the bulk
value of the orientational order parameter.

In order to integrate with respect to the angle in
(\ref{MFA_s}), we should separate the angle $\Omega_{in}$
between the particles and the director and the angle
$\Omega_{nw}$ between the nematic director and the surface. To
this end, we express the Legendre polynomial in the potential
$\nu(r_{12},\Omega_1\Omega_2)$ in terms of spherical harmonics
$Y_{2m}(\Omega)$ as
$P_2(\cos{\Omega_{12}})=({1}/{5})\sum\limits_{m}{Y^{*}_{2m}(\Omega_{1n})Y_{2m}(\Omega_{2n})}.$
As a result,
\begin{align}
\label{MFA_s2}
\frac{\rho(\mathbf{r}_1,\Omega_{1n},\Omega_{wn})}{\rho^\mathrm{bulk}(\Omega_{1n})}
=\exp\left\{-\left[V_0(\mathbf{r}_1,\Omega_{wn})-V_0^\mathrm{b}\right]
-\frac{1}{\sqrt{5}}
\sum\limits_m Y_{2m}(\Omega_{1n})\left[V_{2m}(\mathbf{r}_1,\Omega_{wn})-V_{2m}^\mathrm{b}\right]\right\},
\end{align}
where the mean field potentials
\begin{align}
V_0(\mathbf{r}_1,\Omega_{wn})&=\beta\int\nu_0(r_{12})\rho(\mathbf{r}_2,\Omega_{wn})\rd\mathbf{r}_2\,,\\
V_{2m}(\mathbf{r}_1,\Omega_{wn})&=\beta\int\nu_2(r_{12})S_{2m}(\mathbf{r}_2,\Omega_{wn})\rd\mathbf{r}_2\,.
\end{align}
The bulk values of these quantities are
$V_0^\mathrm{b}=\varkappa_0^2/\alpha_0^2$,
 $V_{20}^\mathrm{b}=\varkappa_2^2 S_\mathrm{b}/\alpha_2^2$,  $V_{2m}^\mathrm{b}=0$
for $m\ne0$, where  $\varkappa_0^2=4\pi\rho_\mathrm{b}\beta A_0$,
\begin{align}
\rho(\mathbf{r},\Omega_{wn})=\int\rho(\mathbf{r},\Omega_{1n},\Omega_{wn})\rd\Omega_{1n}
\end{align}
is the density profile. The property
\begin{align}
\label{shva}
S_{2m}(\mathbf{r},\Omega_{wn})=\frac{1}{\sqrt{5}}\int\rho(\mathbf{r},\Omega_{1n},\Omega_{wn})Y_{2m}(\Omega_{1n})\rd\Omega_{1n}
=\rho(\mathbf{r},\Omega_{wn})\,S_{2m}^*(\mathbf{r},\Omega_{wn}),
\end{align}
where $S_{2m}^*(\mathbf{r},\Omega_{wn})$ are the order
parameter profiles. Far from the wall
$S_{20}^{*}(\mathbf{r},\Omega_{wn})\rightarrow S_\mathrm{b}$,
$S_{2m}^*(\mathbf{r},\Omega_{wn})\rightarrow 0$ for $m\ne0$.

Taking the gradient of equation (\ref{MFA_s2}) we have
\begin{align}
\label{grad}
\frac{1}{\rho(\mathbf{r},\Omega_{1n},\Omega_{wn})}\pmb{\nabla}\rho(\mathbf{r},\Omega_{1n},\Omega_{wn})=
\mathbf{E}_0(\mathbf{r},\Omega_{wn})+\frac{1}{\sqrt{5}}\sum\limits_m Y_{2m}(\Omega_{1n})
\mathbf{E}_{2m}(\mathbf{r},\Omega_{wn}),
\end{align}
where we define an equivalent of the electric field by
\begin{align}
\label{electric}
\mathbf{E}_0({\mathbf{r},\Omega_{wn}})\equiv -\pmb{\nabla} V_0(\mathbf{r},\Omega_{wn}),\qquad
\mathbf{E}_{2m}({\mathbf{r},\Omega_{wn}})\equiv -\pmb{\nabla} V_{2m}(\mathbf{r},\Omega_{wn}).
\end{align}
Due to the properties of the Yukawa potential
\begin{align}
\label{hered}
\left(\Delta-\alpha_0^2\right)V_0(\mathbf{r},\Omega_{wn})&=-4\pi\beta A_0 \rho(\mathbf{r},\Omega_{wn}),\\
\label{hered1}
\left(\Delta-\alpha_2^2\right)V_{2m}(\mathbf{r},\Omega_{wn})&=-4\pi\beta A_2S_{2m}(\mathbf{r},\Omega_{wn}).
\end{align}
Due to the translational invariance parallel
to the wall, the functions considered  depend only on the
distance $z$ to the wall, and replacing  (\ref{hered}) and
(\ref{hered1}) into (\ref{grad}), we obtain
\begin{align}
\label{invariant1}
\frac{{\rd}}{{\rd z}}\left[\vphantom{\frac{\alpha_0^2}{2\varkappa_0^2}}\right.\!\!\!\frac{\rho(z,\Omega_{wn})}{\rho_\mathrm{b}}&+\frac{\alpha_0^2}{2\varkappa_0^2}\,
V_0^2(z,\Omega_{wn})-\frac{1}{2\varkappa_0^2}E_0^2(z,\Omega_{wn})\nonumber\\
&+\sum\limits_m\left.\left(\frac{\alpha_2^2}{2\varkappa_2^2}\,
V_{2m}^2(z,\Omega_{wn})-\frac{1}{2\varkappa_2^2}E_{2m}^2(z,\Omega_{wn})\right)\right]=0.
\end{align}
In the bulk, when $z\rightarrow\infty$, we have
$\rho(z,\Omega_{wn})\rightarrow\rho_\mathrm{b}$,
$E_0(z,\Omega_{wn})\rightarrow 0$,
$E_{2m}(z,\Omega_{wn})\rightarrow 0$,
$V_0(z,\Omega_{wn})\rightarrow V_{0}^\mathrm{b}$,
$V_{20}(z,\Omega_{wn})\rightarrow V_{20}^\mathrm{b}$ and
$V_{2m}(z,\Omega_{wn})\rightarrow 0$ for $m\ne 0$. From
Equation (\ref{invariant1}) we see that the quantity in
brackets is constant regardless of the angle $\Omega_{wn}$
between the director and the wall, and, therefore, it can be
evaluated, for instance, in the bulk as the reduced pressure
$\beta P/\rho_\mathrm{b}$ within MFA \cite{kravtsiv}:
\begin{align}
\label{contactd}
\frac{\beta P}{\rho_\mathrm{b}} = 1+\frac{\varkappa_0^2}{2\alpha_0^2}+\frac{\varkappa_2^2}{2\alpha_2^2}\,S_\mathrm{b}^2\,.
\end{align}
Outside the system, where there are no particles, we have
another invariant which is simply
\begin{align}
\frac{\alpha_0^2}{2\varkappa_0^2}\,
V_0^2(z,\Omega_{wn})-\frac{1}{2\varkappa_0^2}E_0^2(z,\Omega_{wn})+
\sum\limits_m\left[\frac{\alpha_2^2}{2\varkappa_2^2}\,
V_{2m}^2(z,\Omega_{wn})-\frac{1}{2\varkappa_2^2}E_{2m}^2(z,\Omega_{wn})\right],
\end{align}
its value is zero far from the interface and, therefore, it is zero at
the interface as well. From the continuity of the potential and its
derivative due to equation~(\ref{hered}) and (\ref{hered1}), we see
that this is also true at the wall just inside the system ${z}
= 0_{+}$. Thus,
\begin{align}
\label{invariant}
\frac{\rho(0_{+},\Omega_{wn})}{\rho_\mathrm{b}}&+\frac{\alpha_0^2}{2\varkappa_0^2}\,
V_0^2(0_{+},\Omega_{wn})-\frac{1}{2\varkappa_0^2}E_0^2(0_{+},\Omega_{wn})\nonumber\\
&+\sum\limits_m
\left[\frac{\alpha_2^2}{2\varkappa_2^2}\,
V_{2m}^2(0_{+},\Omega_{wn})-\frac{1}{2\varkappa_2^2}E_{2m}^2(0_{+},\Omega_{wn})\right]
=\frac{\rho(0_{+},\Omega_{wn})}{\rho_\mathrm{b}}\,.
\end{align}
Since this quantity is constant, we obtain the so-called contact
theorem
\begin{eqnarray}
\label{theorem}
    {\beta P} = {\rho}(0_{+},\Omega_{wn}).
\end{eqnarray}
We should note that the contact theorem was usually proved for
isotropic fluids near a hard wall \cite{contactth1,contactth2}.
The result obtained here is probably the first verification of
the contact theorem for anisotropic fluids at a hard wall.

From equations (\ref{shva})-(\ref{hered1}) we have a set of six
differential equations for unknown functions
$\rho(\mathbf{r},\Omega_{1n},\Omega_{wn})$,
$S_{2m}(\mathbf{r},\Omega_{wn})$,
$E_0(\mathbf{r},\Omega_{wn})$,
$E_{2m}(\mathbf{r},\Omega_{wn})$,
$V_0(\mathbf{r},\Omega_{wn})$,
$V_{2m}(\mathbf{r},\Omega_{wn})$. We note that in the case when
the director is oriented perpendicular to the wall,
$\Omega_{wn}=0$, the singlet distribution function is axially
symmetric. Consequently, in the equations considered  only the
terms with $m=0$ will be present. In this paper, we will
restrict our further consideration to this special case and
consider the solution of the obtained differential equations in
the linear approximation. Far from the wall
$\rho(\mathbf{r},\Omega)\rightarrow\rho^\mathrm{bulk}(\Omega).$ After
linearization of expression (\ref{MFA_s2})
\begin{align}
\rho{'}(z,\Omega)=\left[E_0(z)+E_{20}(z)P_2(\cos\theta)\right]\rho^\mathrm{bulk}(\Omega)
\end{align}
and we have the following system of equations
\begin{align}
\rho{'}(z)&=\left[E_0(z)+S_\mathrm{b}E_{20}(z)\right]\rho_\mathrm{b}\,,\\
S'_{20}(z)&=\left[E_0(z)S_\mathrm{b}+E_{20}(z)\frac{1}{5}\langle Y_{20}^2\rangle_{\Omega}\right]\rho_\mathrm{b}\,,\\
V_0'(z)&=-E_0(z),\qquad V_{20}'(z)=-E_{20}(z)\,,\\
E_0'(z)&=-\alpha_0^2V_0(z)+(\varkappa_0^2/\rho_\mathrm{b})\,\rho(z)\,,\\
E_{20}'(z)&=-\alpha_2^2V_{20}(z)+(\varkappa_2^2/\rho_\mathrm{b})\,S_{20}(z)\,,
\end{align}
where the prime denotes a derivative by $z$ and $\langle
Y_{20}^{k}\rangle_{\Omega}=(1/\rho_\mathrm{b})\int\limits_0^1
Y_{20}^{k}(\Omega)\rho^\mathrm{bulk}(\Omega)\rd\cos\theta$ ($k=1,2$).
This system reduces to two second order differential equations
for $E_0(z)$ and $E_{20}(z)$
\begin{align}
\label{system_E1d}
E_0''(z)&=E_0(z)\left(\varkappa_0^2+\alpha_0^2\right)+E_{20}(z)\varkappa_0^2\,S_\mathrm{b}\, ,\\
\label{intermsd}
E_{20}''(z)&=E_0(z)\varkappa_2^2\,S_\mathrm{b}
 +E_{20}(z)\left[\alpha_2^2+(\varkappa_2^2/5)\,\langle Y_{20}^2\rangle_{\Omega} \right],
\end{align}
which can be solved with the boundary condition that should
include the contact theorem (\ref{theorem}). Thus,
\begin{align}
\label{dprofile}
\frac{\rho(z)}{\rho_\mathrm{b}}=1&-\frac{\lambda_0^2-\alpha_2^2-\frac{1}{5}\varkappa_2^2
\big(\,\langle Y_{20}^2\rangle_{\Omega}-
\,\langle Y_{20}\rangle_{\Omega}^2\big)}{\varkappa_2^2\,S_\mathrm{b}}\,B_1\,\re^{\displaystyle-\lambda_0 z}\nonumber\\
&-\frac{\lambda_2^2-\alpha_2^2-\frac{1}{5}\varkappa_2^2
\big(\,\langle Y_{20}^2\rangle_{\Omega}-
\,\langle Y_{20}\rangle_{\Omega}^2\big)}{\varkappa_2^2\,S_\mathrm{b}}\,B_2\,\re^{\displaystyle-\lambda_2 z},\\
\frac{S_{20}(z)}{\rho_\mathrm{b}\,S_\mathrm{b}}=1&-\frac{\left(\lambda_0^2-\alpha_2^2\right)}{\varkappa_2^2\,S_\mathrm{b}}\,B_1\,\re^{\displaystyle-\lambda_0 z}
-\frac{\left(\lambda_2^2-\alpha_2^2\right)}{\varkappa_2^2\,S_\mathrm{b}}\,B_2\,\re^{\displaystyle-\lambda_2 z},
\label{sprofile}
\end{align}
where
\begin{align}
\label{lambda0}
\lambda_{0,2}^2=\frac{1}{2}\left[\varkappa_0^2+\alpha_0^2
+\frac{\varkappa_2^2}{5}\langle Y_{20}^2\rangle_{\Omega}+
\alpha_2^2\pm\sqrt{\left(\varkappa_0^2+
\alpha_0^2-\frac{\varkappa_2^2}{5}\langle Y_{20}^2\rangle_{\Omega}-\alpha_2^2\right)^2
+4\varkappa_0^2\varkappa_2^2\,S_\mathrm{b}^2}\right],
\end{align}
\begin{align}
B_1=\frac{\varkappa_2^2\,S_\mathrm{b}}{2\left[\lambda_0^2-\lambda_2^2\right)}\left[-\frac{\varkappa_0^2}{\alpha_0^2}
+\frac{\lambda_2^2-\alpha_2^2-({\varkappa_2^2}/{5})\langle Y_{20}^2\rangle_{\Omega}}{\alpha_2^2}\right],\qquad
B_2=-\frac{\varkappa_2^2\,S_\mathrm{b}}{2\alpha_2^2}-B_1\,.
\end{align}
\begin{figure}
\centerline{
\includegraphics[width=0.47\textwidth]{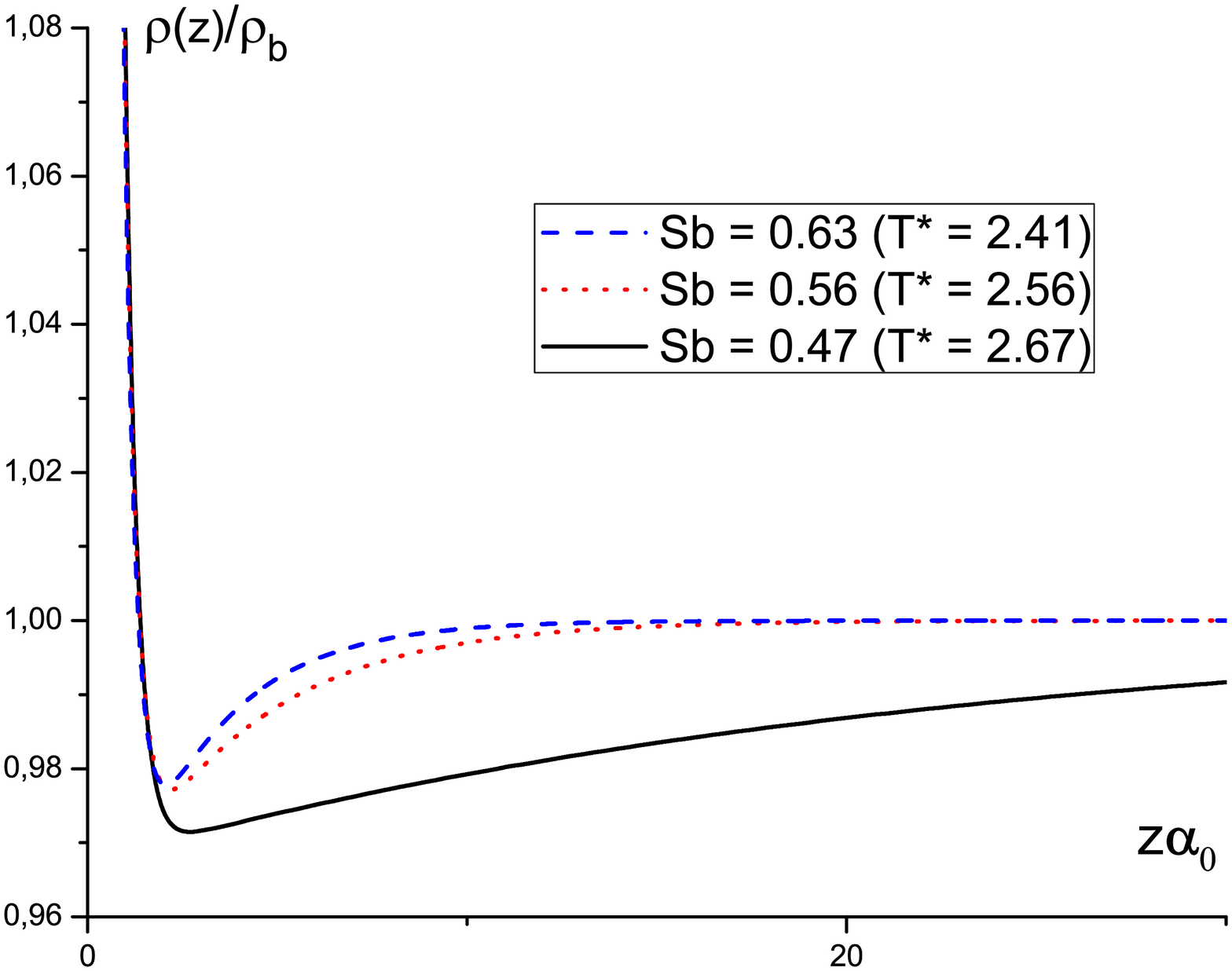} 
\hfill
\includegraphics[width=0.49\textwidth]{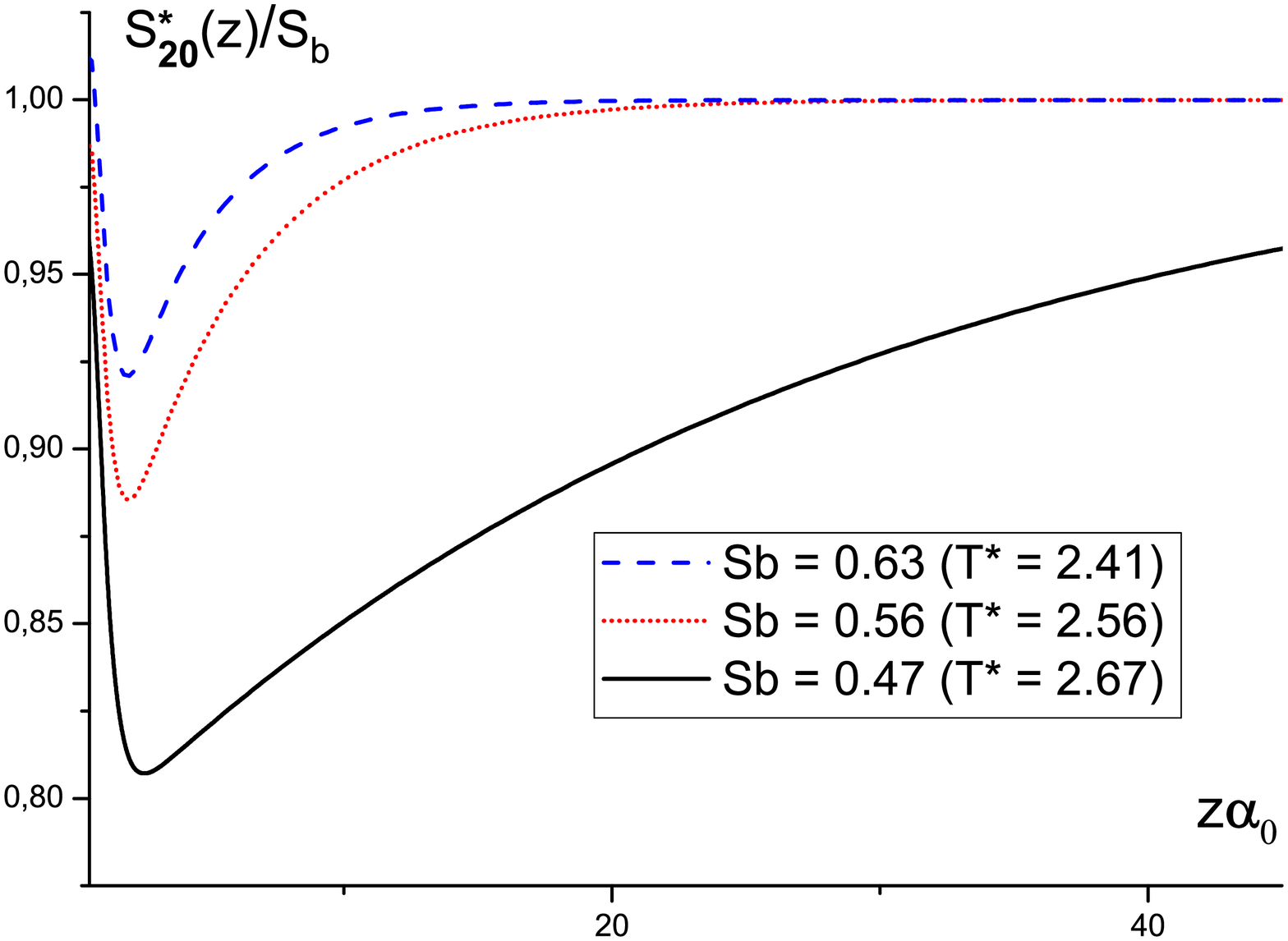}
}
\caption{(Color online) The density and the order parameter profiles in the linearized approximation for $\rho_\mathrm{b}/\alpha_0^3=0.5$, $A_0/A_2=2.2$,
$\alpha_0/\alpha_2=1.25$. Different lines correspond to different values
of $S_\mathrm{b}$ [i.e., the reduced temperature $T^*=1/(\beta A_2\alpha_2)$].}
\label{density_un}
\end{figure}
The values of parameters $\lambda_0$ and $\lambda_2$ coincide
with similar parameters obtained in the bulk case after
including the Gaussian fluctuations \cite{kravtsiv}. Similarly to the bulk case, parameters $\lambda_0$ and $\lambda_2$
characterize the screening of the repulsive isotropic and the
attractive anisotropic interactions, respectively. The density
profile $\rho(z)$ and the order parameter profile
$S_{20}^{*}(z)=S_{20}(z)/\rho(z)$ calculated from equations
(\ref{dprofile})--(\ref{sprofile}) are presented in figure~\ref{density_un}. As
we can see, both profiles have layer-like forms. At the surface,
the fluid is more dense and orientationally ordered than in the
bulk. But at higher distances there is a large region where the
fluid is more diluted and less orientationally ordered than in
the bulk. This result is obtained in the MFA. For isotropic
fluids, as was shown in \cite{molphys}, the inclusion of
fluctuation terms leads to the depletion effect. We can suppose
that a similar effect will take place for the considered
anisotropic fluid. It means that the inclusion of fluctuation terms
does not change our conclusion drawn from figure~\ref{density_un} concerning the
existence of a more diluted and less ordered
region of the fluid  near the surface compared to the bulk region.

Other interesting phenomena that can be observed within the
framework of the formulated MFA approach are connected with the
possibility that the angle between the nematic director and the
surface can change near the surface. Consequently, phases with the order parameters
$S_{2m}^{*}(z)=S_{2m}(z)/\rho(z)$ ($m\ne 0$) can appear near the
surface. These
phenomena will be investigated in a separate paper.

In conclusion, using the field theoretical approach we have
formulated the mean field approximation as the first starting
point for the description of a nematic fluid at a hard wall.
For the first time, an exact derivation of the
contact theorem for anisotropic fluids has been presented within the MFA. It
has been shown that the contact value of the density profile is
determined by the pressure of the fluid in the bulk volume and
does not depend on the angle between the nematic director and
the surface. For the case of the director being oriented
perpendicular to the wall, in the linear approximation for the
MFA, we have obtained analytical expressions for the density and
order parameter profiles. It has been shown that at some values
of the parameters of the model considered, the fluid near the
surface can be more diluted and less orientationally ordered
than in the bulk.

\section*{Acknowledgements}
The authors are grateful for the support to the National
Academy of Sciences of Ukraine and to the Centre National de la
Recherche Scientifique in the framework of the PICS project.

\ukrainianpart

\title{Нематичний плин біля твердої поверхні у наближенні середнього поля}
\author{М. Головко\refaddr{label1}, І. Кравців\refaddr{label1}, Д. ді Капріо\refaddr{label2} }
\addresses{
\addr{label1} Інститут фізики конденсованих систем НАН України,
  вул. І. Свєнціцького, 1, 79011 Львів, Україна
\addr{label2} Лабораторія електрохімії, хімії поверхонь і енергетичного моделювання, відділення хімії вищої національної школи ПаріТех, аб. скринька 39, пл. Жуссю, 4, 75005 Париж, Франція
}

\makeukrtitle

\begin{abstract}
\tolerance=3000
В рамках теоретико-польового підходу вивчається нематогенний плин Майєра-Заупе біля твердої поверхні. У розглядуваній моделі парний потенціал взаємодії складається з ізотропного та анізотропного Юкавівських доданків. У наближенні середнього поля доведено контактно теорему. Для випадку, коли директор направлений перпендикулярно до стінки, отримано аналітичні вирази для профілів густини та параметра порядку. Показано, що у певній термодинамічній області нематичний плин поблизу поверхні може бути більш розрідженим і менш орієнтаційно впорядкованим, ніж в об'ємній області.

\keywords нематичний плин Майєра-Заупе, теоретико-польовий підхід, поверхня розділу, контактна теорема
\end{abstract}


\clearpage

\begin{thebibliography}{99}

\bibitem{jerome}
    Jerome~B.,~Rep.~Prog.~Phys.,~1991,~\textbf{54},~391;
    \bibdoi{10.1088/0034-4885/54/3/002}.
\bibitem{sokol1} Sokolovska~T.G.,~Sokolovskii R.O.,~Patey G.N.,
    Phys. Rev. Lett., 2004, \textbf{92}, 185508; \\ \bibdoi{
    10.1103/PhysRevLett.92.185508}.
\bibitem{sokol2} Sokolovska~T.G.,~Sokolovskii R.O.,~Patey G.N.,
    J. Chem. Phys., 2005, \textbf{122}, 034703; \bibdoi{10.1063/1.1825373}.
\bibitem{hend} Henderson~D., Abraham F.F., Barker J.A., Mol.
    Phys., 1976, \textbf{31}, 1291; \bibdoi{
    10.1080/00268977600101021}.
\bibitem {holsok} Holovko~M., Sokolovska~T., J. Mol. Liq.,
    1999,
     \textbf {82}, 161;
     \bibdoi{10.1016/S0167-7322(99)00098-7}.
\bibitem{contactth1}Henderson D., Blum L., Lebowitz J.L., J.
    Electroanal. Chem., 1979, \textbf{102}, 315;
    \bibdoi{10.1016/S0022-0728(79)80459-3}.

\bibitem{contactth2}Holovko M., Badiali J.P., di Caprio D., J.
    Chem. Phys., 2005, \textbf{123},
    234705;~\bibdoi{10.1063/1.2137707}.
\bibitem{ionic1} Di~Caprio~D., Stafiej~J., Badiali~J.P.,
    Mol. Phys., 2003, \textbf {101},
    2545;~\bibdoi{10.1080/0026897031000154293}.

\bibitem{castaba1}Di~Caprio~D., Stafiej~J., Badiali~J.P.,
    J. Chem. Phys., 1998, \textbf {108},
    8572;~\bibdoi{10.1063/1.476286}.

\bibitem{castaba2}Di~Caprio~D., Stafiej~J., Borkowska Z.,
   J.~Electroanal. Chem., 2005, \textbf{41},
   582;~\bibdoi{10.1016/j.jelechem.2005.02.008}.
\bibitem {molphys} Di~Caprio~D., Stafiej~J., Holovko~M.,
    Kravtsiv~I., Mol. Phys., 2011, \textbf {109},
    695;~\bibdoi{10.1080/00268976.2010.547524}.


\bibitem{masa} Maier~W., Saupe~A.,
    Z.~Naturforsch. A, 1959,
    \textbf{14},
    882.

\bibitem{masa_2} Maier~W., Saupe~A.,
    Z.~Naturforsch. A, 1960, \textbf {15}, 287.

\bibitem{cond}Holovko~M., di~Caprio~D., Kravtsiv~I.,
    Condens. Matter Phys., 2011, \textbf{14},
    33605;~\bibdoi{10.5488/CMP.14.33605}.

\bibitem{kravtsiv} Kravtsiv I., Holovko M., di Caprio D., Mol.
    Phys., 2013 (in press);~\bibdoi{10.1080/00268976.2012.762615}.

\bibitem{gra} Gray~C.G., Gubbins~K.E., Theory of Molecular
    Fluids, Clarendon press, Oxford, 1984.

\end{thebibliography}
\end{document}